\documentclass[useAMS,onecolumn]{mn2e}
\usepackage{amssymb}
\usepackage{graphicx}
\def\apj{ApJ}
\def\mnras{MNRAS}
\def\araa{ARA\&A}
\def\nat{Nature}

\usepackage[dvips]{color}

\begin{document}

\title[An MHD model for episodic jets]
{A Magnetohydrodynamical Model for the Formation of Episodic Jets}
\author[F. Yuan, J. Lin, K. Wu, and L. C. Ho]
{Feng Yuan$^{1,2}$\thanks{E-mail: fyuan@shao.ac.cn (F.Y.),
jlin@ynao.ac.cn (J.L.), kw@mssl.ucl.ac.uk (K.W.), and lho@ociw.edu (L.H.)}, Jun
Lin$^{3,4}$, Kinwah Wu$^{5}$, and Luis C. Ho$^{6}$\\
$^{1}$Shanghai Astronomical Observatory, Shanghai 200030, China;\\
$^{2}$Joint Institute for Galaxy and Cosmology (JOINGC) of SHAO and USTC;\\
$^{3}$National Astronomical Observatories/Yunnan Observatory,
Chinese Academy of Sciences, Kunming 650011, China;\\
$^{4}$Harvard-Smithsonian Center for Astrophysics, 60 Garden Street,
Cambridge, MA 02138, USA;\\
$^{5}$Mullard Space Science Laboratory, University College London,
  Holmbury St Mary, Surrey RH5 6NT, UK;\\
$^{6}$The Observatories of the Carnegie Institution of Washington,
813 Santa Barbara Street, Pasadena, CA 91101, USA; } \maketitle

\begin{abstract}
Episodic ejection of plasma blobs have been observed
in many black hole systems. While steady, continuous jets are believed
to be associated with large-scale open magnetic fields,
what causes the episodic ejection of blobs remains unclear.
Here by analogy with the coronal mass ejection on the Sun,
we propose a magnetohydrodynamical model for episodic ejections from black 
holes associated with the closed magnetic fields in an 
accretion flow. Shear and turbulence of the accretion flow deform the
field and result in the formation of a flux rope in the disk corona.
Energy and helicity are accumulated and stored until a threshold is reached.
The system then loses its equilibrium and the flux rope is thrust outward by 
the magnetic compression force in a catastrophic way.
Our calculations show that for parameters appropriate for the black hole
in our Galactic center, the plasmoid can attain relativistic speeds
in about 35 minutes.
\end{abstract}


\section{Introduction}

Jets are ubiquitous in astrophysical accreting systems. 
Large-scale jets tend to be steady and continuous. 
There are also intermittent, episodic outflows from the accretion systems, 
which are associated with flare emission.  
In Sgr~A*, the massive black hole in the Galactic center,
radio, infrared and X-ray flares occur several times a day (Eckart et al. 2006), 
showing delays among the peaks in the light curves at different 
wavebands (Yusef-Zadeh et al. 2006). 
The delays, together with the fast rise and slow decay in the brightness 
and the polarization of the flare emission, are attributed to 
the ejection and expansion of plasmoids from the accretion flow 
(Yusef-Zadeh et al. 2006; van der Laan 1966). However, despite its close
distance, continuous jets have not been detected in Sgr A* even with the
highest VLBI resolution. X-ray and radio 
monitoring observations of the active galaxy 3C~120 over three years 
also showed episodic ejections, in the form of 
bright superluminal knots (Marscher et al. 2002). 
Such knots in jets are very common in active galactic nuclei (e.g., M~87). 
They are usually explained by the collisions of shells ejected from 
the central engine. It has been proposed that plasmoid 
ejections should be considered transient or ``type II'' 
jets (Fender \& Belloni 2004; Fender, Belloni \& Gallo 2004).

Similar plasmoid ejections have also been observed in Galactic microquasars, 
such as GRS 1915+105 (Mirabel \& Rodriguez 1994; Mirabel et al. 1998; 
Fender et al. 1999), GRO J1655$-$40 (Hjellming \& Rupen 1995), 
and XTE J1550$-$564 (Corbel et al. 2002). 
Brief, intense radio flares, accompanied by X-ray flares  
(and infrared flares as well in GRS 1915+105) are 
observed during the transition from the hard X-ray spectral state 
to the soft X-ray spectral state. 
The flares have been associated with the ejections of plasmoids, 
as in the case of Sgr~A*. The presence of plasmoid ejections 
in Sgr A* is based on interpretation of the 
simultaneous light curves at different wavebands.
By contrast, plasmoid ejecta from microquasars are 
often clearly resolvable in radio images, and they convincingly show 
relativistic motion away from the central core.

The characteristics of episodic plasmoid ejections and steady, continuous jets 
are distinguishable. 
Both kinds of outflows have been observed in the same individual microquasar. 
While episodic ejections occur during the transition from the 
hard to the soft state (Fender, Belloni \& Gallo 2004), continuous jets are seen 
only in the hard state. Observations generally show larger Lorentz 
factor for transient jets (Fender, Belloni \& Gallo 2004). 
Also, their radio spectrum evolves 
rapidly and the emission becomes optically thin, in contrast to 
continuous jets whose bright radio emission remains opaque. 
There is evidence that the emission from ejected plasmoids is more highly 
polarized than the emission from the continuous jets (Fender \& Belloni 2004). 

Magnetohydrodynamical numerical simulations of accretion flows 
assuming a weak initial magnetic field have shown impulsive mass ejections 
embedded in steady, continuous jet-like outflow (Machida, Hayashi \& Matsumoto
2000; De Villiers, Hawley \& Krolik 2003). 
The ejection events are quasi-periodic, and the time interval between 
successive ejections is roughly $1600~GM/c^3$ (De Villiers, Hawley \& Krolik 2003), 
where $G$, $c$ and $M$ are the gravity constant, the speed of light, 
and the black hole mass, respectively. For parameters appropriate 
for Sgr~A*, the time interval is $\sim 6$ hours. If we assume that 
each flare of reasonably high intensity is associated with an ejection event, 
this timescale is consistent with those seen in current 
observations (Yusef-Zadeh et al. 2006). 
  
The formation of the continuous jets has been widely studied and models have 
been proposed (e.g., Blandford \& Znajek 1977; Blandford \& Payne 1982).
However, the origin of episodic jets has remained unclear.
In this paper, by analogy with the coronal mass ejection (CME) phenomena in the Sun,
we propose a magnetohydrodynamical model for episodic jets (\S2). 
We then explain how to understand the above-mentioned various observations 
characteristic of episodic jets based on this model (\S3). Specifically, 
our calculations show that for parameters appropriate for the black hole 
in our Galactic center, the plasmoid can 
attain relativistic speeds in about 35 minutes.

\section{An MHD model for episodic jets}

\begin{figure*} 
\includegraphics[width=16cm]{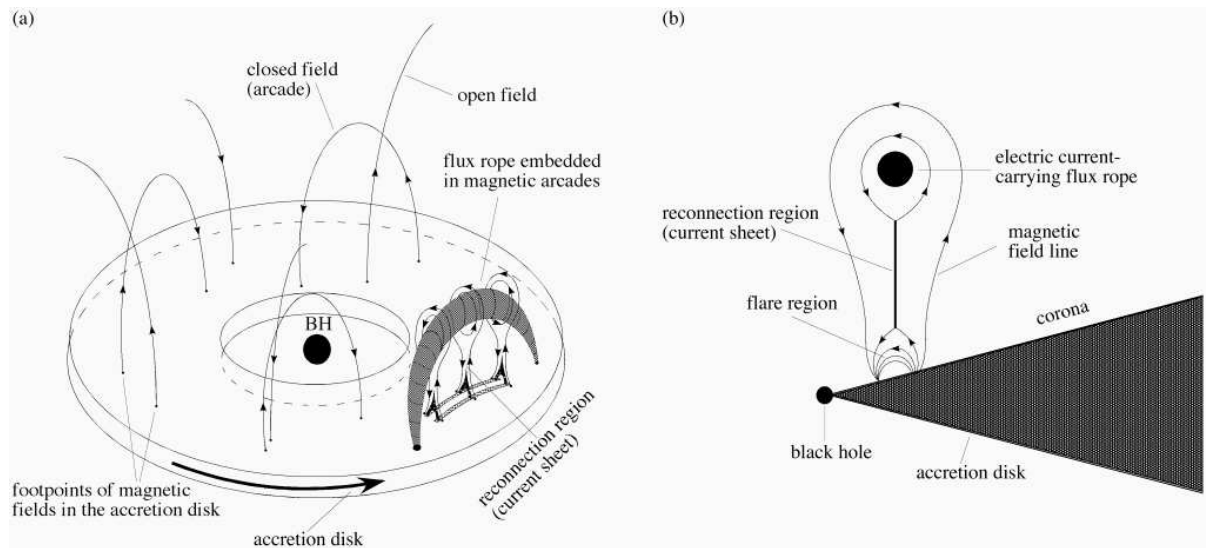}
\caption{Illustration of the formation of a
flux rope (panel a) and the ejection of the flux rope and the
associated flare (panel b). (a) The accretion
flow surrounding the central black hole consists of a main disk body and a
corona envelope. For convenience the disk is drawn to be 
geometrically thin but it actually should be thick (see footnote 2 and the 
right plot of this figure).  Magnetic arcades emerge from the disk into
the corona, and a flux rope is formed, as a result of the motion of their
footpoints and subsequent magnetic reconnection. Magnetic energy and helicity
are continuously transported into the corona and stored in its magnetic field
until the energy exceeds a threshold. (b) The flux rope is then ejected,
forming a current sheet. Magnetic reconnection
occurs in the current sheet, and subsequently the magnetic tension becomes
much weaker than the magnetic compression. This results in the energetic
ejection of the flux rope. The plasma heated in the magnetic reconnection
process produces flares.}

\end{figure*}

We note that two-component magnetic outflows/ejections have been found  
in a variety of astrophysical environments. A well-studied system is 
the outflows of the Sun consisting of fast solar winds and coronal mass 
ejections (CMEs). The fast solar wind is relatively steady, continuous 
and smooth. It originates from the solar surface regions with open magnetic 
field lines. CMEs are, however, episodic, and they are ejected from coronal 
regions with closed magnetic field lines (magnetic arcades). 
The speeds of CMEs can reach up to 2000~${\rm km\,s^{-1}}$ and beyond.  The 
rate of CME occurrence varies from once a few weeks during the solar minimum 
to several times per day at solar maximum (Zhang \& Low 2005; Lin, 
Soon \& Baliunas 2003). 

In the Sun, magnetic arcades generally emerge into the tenuous corona 
from the denser solar photosphere, with their footpoints anchored in the 
photosphere. The configuration of the coronal magnetic field is thought to be 
controlled by convective turbulence in the solar photosphere because of the 
freezing of the field in the plasma. Convective turbulence motion in the 
photospheric plasma leads to the formation of coronal flux ropes,   
manifested as dark filaments and bright prominences in observations 
(Zhang \& Low 2005). The ropes are in an equilibrium configuration 
when there is a balance between 
the forces due to magnetic compression and magnetic tension from below and 
above the flux rope (refer to Figure 1a). 
Nevertheless, the equilibrium is temporary. The turbulence in the 
photosphere inevitably cause a build-up of stress and helicity. 
They also convert the kinetic energy in the photosphere into the 
magnetic energy in the corona (Zhang \& Low 2005; Lin, Soon \& Baliunas 2003).
When the energy accumulation exceeds a certain threshold, 
the confinement in the magnetic arcade breaks down, and the flux rope 
gets thrust outward in a catastrophic manner (Forbes \& Isenberg 1991; 
Lin \& Forbes 2000; Lin, Soon \& Baliunas 2003)\footnote{
How to determine the timescale of the above energy accumulation process
is still an open question. It should be related to
the energy transfer speed from the photosphere to the corona, which is determined by
the Alfv\'{e}n timescale, and the value of the threshold energy.}
Then the magnetic field is severely stretched and a neutral 
region --- the current sheet --- develops, 
separating magnetic fields of opposite polarity (refer to Figure 1b).  
Dissipation, facilitated by microscopic plasma instabilities, 
leads to rapid magnetic reconnection in the current sheet (Lin \& Forbes 2000; 
Lin, Soon \& Baliunas 2003). This then greatly relaxes the magnetic
tension and helps the compression push
the rope through the corona smoothly, developing into a CME.
On the one hand, magnetic reconnection converts the stored magnetic 
energy into the microscopic particle kinetic energy in the plasma,  
which ignites the radiative flares; on the other hand, it transfers the 
magnetic energy into bulk kinetic energy that propels CME 
propagation (Zhang \& Low 2005; Lin \& Forbes 2000). The timescale of the
above process of magnetic energy release is determined by the local Alfv\'{e}n one. 

The similar morphology and characteristics between the outflow components
in accreting black holes and in the Sun indicates the operation of a 
common physical mechanism. We therefore seek to build a model for the 
magnetic outflows in accreting 
black holes in light of current understanding of magnetic outflows in 
the solar environment. In current models for continuous jets,
a large-scale open magnetic field is required to extract energy from
the rotation of the black hole or the flow in the accretion 
disk (Blandford \& Znajek 1977; Blandford \& Payne 1982). 
If continuous jets correspond to the smooth 
solar wind component, we propose that the transient, type-II jets are 
analogous to CMEs, which result from the disruption of closed field 
lines in the corona above the accretion disk.

In addition to observational evidence mentioned above, there are 
compelling theoretical 
reasons to believe that similar physical processes operate in accreting 
black hole and in the solar environment. Numerical simulations have shown 
that the structure of a hot accretion flow\footnote{ 
In this paper we focus on hot accretion flows. A well-known example 
of this type of accretion model is advection-dominated accretion flows (ADAFs) or
radiatively inefficient accretion flow (RIAFs) (e.g.,
Narayan \& Yi 1994,1995; Yuan, Quataert \& Narayan 2003).
Compared to the standard optically thick, geometrically thin disk 
(Shakura \& Sunyaev 1973), a hot accretion flow
is geometrically thick, hot, and optically thin. See Narayan, Mahadevan 
\& Quataert 1998 for a review of ADAFs.} is
very similar to the solar atmosphere --- a dense disk 
enveloped by a tenuous corona (Machida, Hayashi \& Matsumoto 2000;
De Villiers, Hawley \& Krolik 2003; De Villiers et al. 2005).
An accretion disk with 
angular momentum transport regulated by magneto-rotational instability
is intrinsically turbulent (Balbus \& Hawley 1998). Loops of magnetic field emerge 
into the disk corona. Since their foot points are anchored in the accretion 
flow which is differentially rotating and turbulent, reconnections and 
flares occur subsequently (Romanova et al. 1998;
Blandford 2002; Hirose et al. 2004; Machida, Nakamura
\& Matsumoto 2004; Uzdensky \& Goodman 2008; Goodman \& Uzdensky 2008). 
Especially, Uzdensky \& Goodman (2008) recently present
a physics-based statistical theory governing the evolution of flux loop,
considering their emergence and reconnection. 
We believe that this mechanism gives rise to the 
flares in Sgr~A* that are not accompanied by a CME-type event. 

\begin{figure*}
\vspace{-2.in}
\includegraphics[width=16cm]{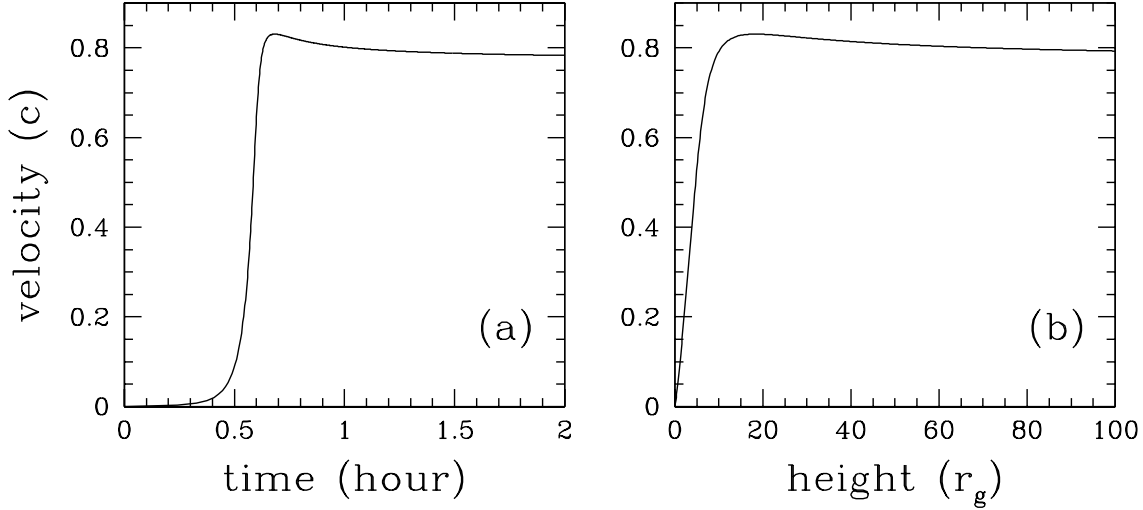}\vspace{-1.4in}
\caption{The calculated velocity of the ejected plasmoid from
Sgr A* as a function of time (a) (in unit of hour) and height (b) (in unit of 
$r_g$). For comparison, the light-crossing time across $r_g$ 
is $\sim 20$ seconds. The flux rope does not
gain the kinetic energy immediately after the catastrophe. Instead it
takes about 25 minutes to accelerate from rest to $\sim$0.02c because of
its inertia and the weakness of the magnetic compression force at the beginning.
In the following 10 minutes, the catastrophe together with magnetic
reconnection rapidly energize the flux rope and accelerate it to $\sim$0.8c.}
\end{figure*}

We argue that the above scenario can develop into a more catastrophic situation.
The reconnection process changes the magnetic field topology (see Figure 1).  
It also redistributes the helicity and stores most of it in a flux rope 
floating in the disk corona, resembling what happens in the Sun 
(Zhang \& Low 2005). The magnetic configuration continues to evolve, after 
the formation of the flux rope. The system eventually loses the equilibrium, 
rapidly expelling the flux rope, and a long current sheet develops behind the 
break-away flux rope (see Forbes \& Isenberg 1991; Lin \& Forbes 2000 and
Figure 1b). Magnetic reconnection then occurs in the sheet, propelling the 
flux rope away from the accretion disk, giving rise to the accretion disk CME. 
This process can occur at any radius, but should be much stronger
in the innermost region where the magnetic field is much stronger
and the emergence of magnetic loops is much more frequent. 
The process also causes intensive heating in the accretion disk 
and its corona (Lin, Soon \& Baliunas 2003; Lin \& Forbes 2000). 
The radiations from the impulsively heated plasma in the accretion disk 
and corona will appear as flares in the observations. Such a scenario 
implies that flares are not purely radiative but also dynamical, and 
provides a very natural explanation to the flares that are often 
associated with mass ejection events, as observed in 
microquasars and Sgr A*. 

\section{Interpreting various observations characteristic of episodic jets}

We now elaborate how plasmoids (flux ropes) from the 
accretion disk corona evolve and propagate in the framework 
of the CME catastrophe model (Lin \& Forbes 2000). 
We use Sgr~A* as an illustration,
as its accretion flow is relatively well understood, 
providing us with reliable estimates of the density, temperature and  
magnetic field in the accretion flow (Yuan, Quataert \& Narayan 2003).  
Consider a plasmoid confined in a coronal magnetic field. 
It is subject to gravity and magnetic forces (Lin, Mancuso \& Vourlidas 2006); 
gas pressure can be neglected since it is assumed to be small compared to
the magnetic pressure. The plasmoid is in a critical position, 
where the loss of equilibrium starts occurring and the plasmoid 
will be thrust outward. After the catastrophic loss of equilibrium, 
the upward motion of the flux rope is governed to the first order
of approximation by:
\begin{equation}
   m \frac{d^{2} h}{dt^{2}}=\frac{1}{c}\left\vert  {\mathbf I}\times 
{\mathbf B}_{\rm ext} \right\vert_{h} - F_{g} \ ,
\end{equation}
where $m$ is the total mass inside the flux rope per unit length, 
$h$ is the height of the flux rope from the surface of the accretion disk, 
$F_{g}$ is the gravitational force, $\mathbf I$ is the integration of the 
electric current intensity $\mathbf j$ inside the flux rope, 
and $B_{\rm ext}$ is the total magnetic field from all the sources 
except $\mathbf I$, 
which includes those inside the disk, on the disk surface, and the 
current sheet. Both the current and magnetic field within and outside of 
the flux rope satisfy the following conditions to zeroth-order
approximation (see Lin \& Forbes 2000 for details): 
\begin{equation}
\mathbf j \times \mathbf B =0 \ ,
\end{equation}
\begin{equation}
\mathbf j=\frac{c}{4\pi} \nabla \times \mathbf B.
\end{equation}

We choose $(x,y)$ coordinate with $y=0$ being the equatorial plane 
and $x=0$ being the rotation axis of the accretion flow \footnote{It 
has been shown (e.g., Uzdensky 2002) that in Cartesian coordinate system
shearing magnetic arcades alone never drives the system to lose 
its equilibrium. This result holds when the magnetic configuration is
``simple''. However, when circular field lines are introduced as in the present 
work, the system can lose equlibrium and develop eruption even in 
Cartesian coordinate system (e.g., see Miki\'{c} et al. 1988 and 
Miki\'{c} \& Linker 1994).}. Suppose that the flux rope is initially 
located at $(x,y)=(5r_{\rm g},10r_{\rm g})$, where 
$r_g\equiv GM/c^{2}=5.9 \times 10^{11} {\rm cm}$ is the gravitational radius
of the black hole in Sgr~A*. Based on studies of 
the accretion flow and its corona in Sgr~A* (Yuan, Quataert \& Narayan 2003;
De Villiers, Hawley \& Krolik 2003), 
we set the density of the rope to $n_0=10^{5} {\rm cm}^{-3}$ 
and the magnetic field to $B_0=16$~G. {These numbers imply 
that the corona is magnetically dominated, and so it is force-free to zeroth-order.} 
The rate of magnetic reconnection, 
namely the Alfv\'{e}n Mach number $M_{A}$, is taken as 0.1 here. It is the 
reconnection inflow speed compared to the local Alfv\'{e}n speed near the 
current sheet.
  
The distribution of the density and magnetic field of the corona 
along the vertical height are very uncertain. Without losing generality, 
we boldly assume that they are similar to those in the solar atmosphere,  
which allows us to directly adopt the well-studied models 
for the solar environment (Lin, Mancuso \& Vourlidas 2006). 
This assumption can be justified because the results do not change significantly 
unless the Alfv\'{e}n speed decreases very rapidly with height 
within the region of interest, which is unlikely given the 
planar configuration of the disk corona. Solving equation (1) yields the 
velocity evolution of the flux rope as shown by Figure 2 (see 
Appendix for details). Our calculations show the flux rope can be 
easily accelerated to a speed of 0.8$c$ in about 35 minutes. 

We now assess the CME scenario using four observational 
characteristics of episodic transient jets. 

{\it (i) Hard to soft X-ray spectral state transition in microquasars.}  
The hard state of microquasars is characterized by a steady hot accretion flow. 
During the transition from the hard state to the soft state, 
the accretion flow changes rapidly from a hot phase to a cold phase. 
The collapse of the accretion flow may lead to a temporal eclipse 
of the continuous jet (Livio, Ogilvie \& Pringle 1999; Fender,
Belloni \& Gallo 2004). The magnetic field in the cold
disk is strongly amplified because of the conservation of the 
magnetic flux during the collapse process. Such a field configuration is highly 
unstable, and as a consequence substantial magnetic flux will be 
expelled out from the disk (Pringle, Rees, \& Pacholczyk 1973; 
Shibata, Tajima \& Matsumoto 1990). 
The readjustment of the field provides the free energy to 
drive ``CME" and power the radiative flares. This naturally explains 
the observed ejection of plasmoids and the radio and X-ray flares 
observed in the hard to soft state transition in microquasars (Fender,
Belloni \& Gallo 2004). 

{\it (ii) Large Lorentz factor.} 
During the disk collapse, a large amount of magnetic energy is liberated on 
a shorter timescale. The acceleration of the plasmoid ejecta is facilitated by 
this impulsive energy release, which is in contrast to the smooth injection of 
energy by an ordinary, steady hot accretion flow that launches the continuous 
jet. It is not surprising that the ejected plasmoids should have a larger 
Lorentz factor than the continuous jet.

{\it (iii) Optically thin emission and strong polarization.} 
Particles are accelerated in the magnetic reconnection current sheet 
as well as in shocks formed when the high-speed plasmoids 
pass through the interstellar medium.  
The radio flares are due to synchrotron radiation 
from the accelerated relativistic electrons; the X-ray flares are 
probably caused by bremsstrahlung, synchrotron-self-Comptonization, or also 
synchrotron radiation. Unlike a normal continuous jet, the plasmoid 
ejecta are single blobs. They expand almost adiabatically 
after leaving the accretion disk and can quickly become 
optically thin in the radio band. The high degree 
of polarization is simply a consequence of the presence of a relatively 
ordered magnetic field enclosing the ejecta and the small optical 
depth in the substantially inflated plasmoid blobs. 

{\it (iv) Presence of bright knots.} 
The accretion disk corona is threaded by both open and closed magnetic 
fields, allowing the coexistence of continuous and transient type II jets. The 
interaction of the very high-speed ejected plasmoid blobs with the slow 
preexisting continuous jet could easily lead to shock formation, which will 
appear as bright knots embedded in steady continuous jet.

\section{ACKNOWLEDGMENTS}

We thank R. Matsumoto, J. E. Pringle, D. Uzdensky, F. Yusef-Zadeh, and 
especially the anonymous referee for 
helpful comments and encouragements. This work was 
supported by the Natural Science Foundation of China (grants 
10773024, 10833002, 10821302, 10825314, 10873030, and 40636031),
One-Hundred-Talent Program of CAS, and the National Basic
Research Program of China (grants 2009CB824800 and 2006CB806303). 
J.L. was supported also by the Chinese Academy of Sciences 
(grant number KJCX2-YW-T04) and NASA (grant number NNX07AL72G) when 
visiting CfA.


\begin{appendix}

\section{Solving Equation (1)}

In this section we describe how we solve equation (1). There are five
parameters involved in the description of the flux rope motion. They 
include the height, the velocity of the flux rope $h$, $\dot{h}$, 
heights of the lower and higher tips of the current sheet $p$ and $q$, 
and the total mass inside the flux rope $m$. (All these heights are 
measured from the disk surface to the zeroth-order approximation in our 
calculations. See also Figure 1b.) In addition to equation (1), we 
need another four equations to close the governing system. Deducing these 
equations is tedious and complicated. We do not duplicate the details 
here and only briefly summarize the results that we are using and the 
related physics. Readers are referred to our previous works for details 
(Lin \& Forbes 2000; Lin, Mancuso \& Vourlidas 2006).

Two terms on the right-hand of equation (1) result from the 
electromagnetic interaction and the gravity of the black hole acting on 
the flux rope, which read as
\begin{eqnarray}
F_{m} &=& \frac{B_{0}^{2}\lambda^{4}}
{8hL_{PQ}^{2}}\left[\frac{H_{PQ}^{2}}
{2h^{2}}-\frac{(\lambda^{2}+p^{2})(h^{2}-q^{2})
}{\lambda^{2}+h^{2}} - 
\frac{(\lambda^{2}+q^{2})(h^{2}-p^{2})}{\lambda^{2}+h^{2}}
\right],
\label{eq:XF1}\\
F_{g} &=& \frac{mg_{0}}{(1+h/\lambda)^{2}}, \label{eq:gravity}
\end{eqnarray}
where $L_{PQ}^{2}=(\lambda^{2}+p^{2})(\lambda^{2}+q^{2})$, 
$H_{PQ}^{2}=(h^{2}-p^{2})(h^{2}-q^{2})$, $B_{0}$ is the average 
field strength, $\lambda$ is the distance between the two footpoints 
anchored on the disk surface, and $g_{0}$ is the gravity near the disk 
surface in the zeroth-order approximation. Substituting equations 
(\ref{eq:XF1}) and (\ref{eq:gravity}) into (1) gives the equation for 
$d\dot{h}/dt=d^{2}h/dt^{2}$, and the equation for $h$ is simply $dh/dt=\dot{h}$.

The first term in the square brackets on the right-hand side of equation 
(\ref{eq:XF1}) denotes the magnetic compression force, which results 
from the interaction of the body electric current in the flux rope with 
the surface current on the disk that is induced by the body current in 
the rope. It is this force that pushes the flux rope outwards (upwards), 
and makes the catastrophic loss of equilibrium in the system of interest 
possible. The other two terms in the brackets come from the interaction 
of the flux rope body current with those inside the current sheet and the 
accretion disk (not including the surface current). The resulting force, 
also known as the magnetic tension, tends to pull the flux rope backwards 
(downwards). With the loss of equilibrium in the system occurring, the 
compression dominates the tension. So the flux rope is thrust outward in a 
catastrophic fashion.

The equations governing the evolution of the current sheet, $dp/dt$ and 
$dq/dt$, 
were deduced from two conditions: one for the current sheet and another 
for the flux rope. The former results from the force-free condition 
outside the current sheet that forces the magnetic field to form a Y-type 
neutral point at each end of the current sheet (see the feature of the 
magnetic field near the two tips of the current sheet shown in Figure 1b), 
and the latter is due to the frozen magnetic flux condition on the 
surface of the flux rope that requires that the total magnetic flux in 
the space between the flux rope surface and infinity remain unchanged 
(e.g., see Forbes \& Isenberg 1991; Lin \& Forbes 2000).

Combining these two conditions yields
\begin{eqnarray}
\frac{dp}{dt} &=& p^{\prime}\dot{h}, \nonumber\\
\frac{dq}{dt} &=& q^{\prime}\dot{h}, \label{eq:eqs_pq}
\end{eqnarray}
where $p^{\prime} = dp/dh$, $q^{\prime} = dq/dh$, and
\begin{eqnarray}
p^{\prime}&=&\frac{\tilde{A_{0h}}A_{Rq}-A_{Rh}A_{0q}} {A_{Rp}A_{0q}-
A_{0p}A_{Rq}},\nonumber\\
q^{\prime}&=&\frac{A_{Rh}A_{0p}-\tilde{A_{0h}}A_{Rp}} {A_{Rp}A_{0q}-
A_{0p}A_{Rq}}, \label{eq:simeqs}
\end{eqnarray}  
with
\begin{eqnarray*}
\tilde{A_{0h}}=\frac{M_{\mbox{\tiny A}}B_{y}^{2}(0,\mbox{ }y_{0})}
{B_{0}\lambda\dot{h}\sqrt{4\pi\rho(y_{0})}}-A_{0h},
\end{eqnarray*}
where $y_{0}=(p+q)/2$ is the height of the current sheet center, 
$M_{A}$ is the reconnection Alfv\'{e}n Mach number, which is the
reconnection inflow speed compared to the local Alfv\'{e}n speed near 
the current sheet, $B_{y}$ and $\rho(y)$ will be given shortly, and 
\begin{eqnarray}
A_{0p}&=&\frac{\lambda p(h^{2}+\lambda^{2})(\lambda^{2}+q^{2})}
{h^{2}q[(\lambda^{2}+p^{2})(\lambda^{2}+q^{2})]^{3/2}}
\left[(h^{2}-q^{2})\Pi\left(\frac{p^{2}}{h^{2}},\frac{p}{q}\right)
-h^{2}K\left(\frac{p}{q}\right)\right]\nonumber\\
A_{0q}&=&\frac{\lambda (h^{2}+\lambda^{2})(\lambda^{2}+p^{2})}
{h^{2}[(\lambda^{2}+p^{2})(\lambda^{2}+q^{2})]^{3/2}}
\left[(h^{2}-q^{2})\Pi\left(\frac{p^{2}}{h^{2}},\frac{p}{q}\right)
-h^{2}K\left(\frac{p}{q}\right)\right]\nonumber\\
A_{0h}&=&-\frac{\lambda}
{h^{3}q\sqrt{(\lambda^{2}+p^{2})(\lambda^{2}+q^{2})}}
\left[h^{2}q^{2}E\left(\frac{p}{q}\right)-
h^{2}(h^{2}+q^{2})K\left(\frac{p}{q}\right)
\right.\nonumber\\&+& \left.
 (h^{4}-p^{2}q^{2})
\Pi\left(\frac{p^{2}}{h^{2}},\frac{p}{q}\right)\right], \label{eq:A0pqh}
\end{eqnarray}
and
\begin{eqnarray}
A_{Rp}
&=&\frac{\lambda p(h^{2}+\lambda^{2})}{q(\lambda^{2}+p^{2})^{2}} \sqrt{
\frac{\lambda^{2}+p^{2}}{\lambda^{2}+q^{2}}} \left\langle\left(1-
\frac{p^{2}}{h^{2}}\right)\Pi\left[\sin^{-1}\left(\frac{q}{h}\right),
\frac{p^{2}}{h^{2}},\frac{p}{q}\right] \right. \nonumber \\
&-& \left. F\left[\sin^{-1}\left(\frac{q}{h}\right), \frac{p}{q}\right]
-\frac{q}{2h}\sqrt{\frac{h^{2}-q^{2}}{h^{2}-p^{2}}} \left\{1 + \ln
\left[\frac{\lambda H_{PQ}^{3}}{r_{00}L_{PQ}(h^{4}-p^{2}q^{2})}\right]
\right\}\right\rangle,\nonumber\\
A_{Rq}
&=&\frac{\lambda (h^{2}+\lambda^{2})}{(\lambda^{2}+q^{2})^{2}} \sqrt{
\frac{\lambda^{2}+q^{2}}{\lambda^{2}+p^{2}}}\left\langle\left(1-
\frac{p^{2}}{h^{2}}\right)\Pi\left[\sin^{-1}\left(\frac{q}{h}\right),
\frac{p^{2}}{h^{2}},\frac{p}{q}\right]\right.\nonumber\\
&-& \left. F\left[\sin^{-1}\left(\frac{q}{h} \right), \frac{p}{q}\right]
-\frac{q}{2h}\sqrt{\frac{h^{2}-p^{2}}{h^{2}-q^{2}}} \left\{1+\ln
\left[\frac{\lambda H_{PQ}^{3}}{r_{00}L_{PQ}(h^{4}-p^{2}q^{2})}\right]
\right\}\right\rangle,\nonumber\\
A_{Rh}
&=&\frac{\lambda}{2h^{2}L_{PQ}H_{PQ}}\left\{\frac{2h^{6}- 2(\lambda
pq)^{2}} {h^{2}+\lambda^{2}} -
\frac{h^{2}(p^{2}+q^{2})(h^{2}-\lambda^{2})}{h^{2}+\lambda^{2}} \right.
\nonumber\\
&+&\left.
(h^{4}-p^{2}q^{2})\ln\left[\frac{\lambda H_{PQ}^{3}}
{r_{00}L_{PQ}(h^{4}-p^{2}q^{2})}\right]\right\} \nonumber\\
&+&\frac{\lambda}{hqL_{PQ}}\left\{(h^{2}+q^{2}) F\left[\sin^{-1}
\left(\frac{q}{h}\right),\frac{p}{q}\right]-q^{2}E\left[\sin^{-1}
\left(\frac{q}{h}\right),\frac{p}{q}\right] \right.\nonumber\\  
&-&\left.
\left(h^{2}-\frac{p^{2}q^{2}}{h^{2}}\right)
\Pi\left[\sin^{-1}\left(\frac{q}{h}\right),\frac{p^{2}}{h^{2}},
\frac{p}{q}\right]\right\}. \label{eq:PDAR}
\end{eqnarray}
Here $K$, $E$, and $\Pi$ in (\ref{eq:A0pqh}) are first, second, and third
kinds of complete elliptic integrals, respectively; $F$, $E$, and $\Pi$
in (\ref{eq:PDAR}) are first, second, and third kinds of incomplete
elliptic integrals, respectively. $B_{y}$ in the expression for 
$\tilde{A_{0h}}$ reads as
\begin{equation}
B_{y}(0, y) = \frac{B_{0}\lambda^{2}(h^{2}+\lambda^{2})
\sqrt{(y^{2}-p^{2})(q^{2}-y^{2})}}
{(y^{2}+\lambda^{2})(h^{2}-y^{2})
\sqrt{(p^{2}+\lambda^{2})(q^{2}+\lambda^{2})}}
\label{eq:By}
\end{equation}
for $p\le y \le q$. Therefore, equations in (\ref{eq:eqs_pq}), together 
with equations (\ref{eq:simeqs}), (\ref{eq:A0pqh}), and (\ref{eq:PDAR}), 
govern the evolution of the current sheet in the eruption.

Lin \& Forbes (2000) pointed out that a fast magnetic reconnection in the 
current sheet is needed in order to convert the magnetic energy into 
heating and the kinetic energy at a reasonable rate so that the evolution in 
the system following the catastrophe is able to develop to a plausible 
eruption. Lin, Mancuso \& Vourlidas (2006) noticed as well that plenty of 
the plasma in the corona is also brought into the current sheet by the 
reconnection inflow, and eventually sent into the rope by the reconnection 
outflow. So the total $m$ inside the rope increases with time. The equation 
governing such a change reads as (Lin, Mancuso \& Vourlidas 2006):
\begin{equation}
\frac{dm}{dt}=B_{0}M_{A}\sqrt{\frac{\rho_{0}}{\pi}}
\frac{\lambda^{2}(q-p)(h^{2}+\lambda^{2})}{(h^{2}-y_{0}^{2})
(y_{0}^{2}+\lambda^{2})}\sqrt{\frac{f(y_{0})(q^{2}-y_{0}^{2})
(y_{0}^{2}-p^{2})}{(p^{2}+\lambda^{2})(q^{2}+\lambda^{2})}},
\label{eq:dmdt2}
\end{equation}
where $f(y)$ is a dimensionless function for the plasma density 
distribution in the corona, and is related to the mass density distribution 
$\rho(y)$ in the expression for $\tilde{A_{0h}}$ such that 
$\rho(y)=\rho_{0}f(y)$, with $\rho_{0}$ being the mass density on the 
surface of the disk.

There is no existing model for the density distribution of a magnetic 
corona above an accretion disk. If we assume that the physical processes 
in the accretion disk corona are similar to that of the Sun, and that the 
dynamical interface between the differentially rotating disk is analogous 
to that of the corona-photosphere interface of the Sun, we may adopt the 
solar model, in which $f(y)$ takes the form (e.g., see Lin, Mancuso \& 
Vourlidas 2006 and references therein):
\begin{equation}
f(y)=a_{1}z^{2}(y)e^{a_{2}z(y)}[1+a_{3}z(y)+a_{4}z^{2}(y)+a_{5}z^{3}(y)],
\label{eq:rho}
\end{equation}
where
\begin{center}
        \begin{tabular}{ll}
$z(y)=\lambda/(\lambda+y)$, & $a_{1}=0.001272$, \\
$a_{2}=4.8039$, & $a_{3}=0.29696$, \\
$a_{4}=-7.1743$, & $a_{5}=12.321$,
        \end{tabular}
\end{center}
with $f(0)=1$.

At this point, we are ready to solve equations (1), (\ref{eq:eqs_pq}), 
(\ref{eq:dmdt2}), and $dh/dt=\dot{h}$ with initial conditions $h(t=0)$, 
$\dot{h}(t=0)$, $p(t=0)$, $q(t=0)$, and $m(t=0)$ being given. We assume 
$h(t=0)=10~r_g$ where the flux rope starts losing its equilibrium and the 
eruption commences, $\dot{h}(t=0)=10^{-5}c$ as an initial perturbation 
velocity, $p(t=0)=q(t=0)=0$, and $m(t=0)=m_{0}$. The results are shown in 
Figure 2.

Our model has adopted certain approximations to simplify the calculations. 
We think this is sufficient for illustrating how plasmoid ejecta can be 
expelled from an accretion  disk and are accelerated to relativistic speeds. 
Note that the  gravitational potential is taken to depend only on the flux 
rope height $h$, although it should be a function of radius 
$r(\equiv \sqrt{x^2+y^2})$. This approximation should not affect our results
since the gravitational force is not important compared to the magnetic 
force. Another simplification is that relativistic effects are not included 
explicitly in equation (1). Special relativity will modify the inertial term,
and general relativity will modify the gravity term. While a relativistic version of
equation (1) should be used in future work, even without 
invoking a fully relativistic treatment, we can see from equation (1) that 
the speed of the plasmoid ejecta would approach the speed of light, provided 
the acceleration region is sufficiently close to the black hole event horizon, 
as the Lorentz force provided by the flux loop needs to overpower gravity in
order to propel the plasmoid ejecta to escape to infinity.

\end{appendix}


\clearpage

\end{document}